\documentclass[aps,prl,twocolumn]{revtex4}
\usepackage{epsfig}
\usepackage[dvipsnames,usenames]{color}
\usepackage{color}
\usepackage{amsmath}
\usepackage{amssymb}

\emergencystretch=\maxdimen
\newcommand{\bK}{{\bf K}}

\begin{document}
\title{$d$-wave superconductivity in the presence of a near neighbor
Coulomb
repulsion}
\author{M. Jiang}
\affiliation{Institute for Theoretical Physics, ETH Zurich, Switzerland}
\email{nonlinearf1@gmail.com}
\author{U.R. H\"{a}hner}
\affiliation{Institute for Theoretical Physics, ETH Zurich, Switzerland}
\author{T.C. Schulthess}
\affiliation{Institute for Theoretical Physics, ETH Zurich, Switzerland}
\author{T.A. Maier}
\affiliation{Computational Science and Engineering Division and Center for
Nanophase Materials Sciences, Oak Ridge National Laboratory, USA}
\email{maierta@ornl.gov}

\begin{abstract}
Dynamic cluster quantum Monte Carlo calculations for a doped two-dimensional
extended Hubbard model are used to study the stability and dynamics of
$d$-wave pairing when a near neighbor Coulomb repulsion $V$ is present in
addition to the on-site Coulomb repulsion $U$. We find that $d$-wave pairing
and the superconducting transition temperature $T_c$ are only weakly
suppressed as long as $V$ does not exceed $U/2$. This stability is traced to
the strongly retarded nature of pairing that allows the $d$-wave pairs to
minimize the repulsive effect of $V$. When $V$ approaches $U/2$, large
momentum charge fluctuations are found to become important and to give rise to
a more rapid suppression of $d$-wave pairing and $T_c$ than for smaller $V$.
\end{abstract}

\maketitle


In conventional superconductors, the retardation of the electron-phonon
pairing interaction is essential to overcome the Coulomb repulsion between
electrons and to give a net attractive interaction \cite{AndersonMorel1962}.
In strongly correlated superconductors, such as the cuprates, heavy fermion or
iron-based materials, in contrast, it is a sign-change in the pair wave
function that allows the Cooper pairs to minimize the repulsive effect of the
strong local Coulomb repulsion \cite{ScalapinoRMP}. For example, the $d_
{x^2-y^2}$-wave pair state in the cuprates completely avoids the local Coulomb
repulsion because of the sign change under 90 degree rotation and the related
lack of a local amplitude.

However, in realistic systems, the Coulomb repulsion is hardly screened to a
completely local interaction, but has a short-ranged non-local contribution.
For the cuprates, S\'en\'echal {\it et al.} \cite{Tremblay2013} and Reymbaut
{\it et al.} \cite{Tremblay2016} estimated a near neighbor Coulomb repulsion
of $\sim 400$ meV. If the Cooper pairs are made up of electrons sitting on
neighboring sites, such as in the $d_{x^2-y^2}$-wave state, this non-local
repulsion is expected to have detrimental effects on the pairing. This raises
the important question of how much the superconducting transition temperature
$T_c$ will be reduced by a non-local Coulomb repulsion and whether retardation
effects, similar to the case of electron-phonon mediated pairing, can play a
role in stabilizing superconductivity in the presence of a non-local
repulsion.

Here we examine these questions in a 2D extended Hubbard model. Its
Hamiltonian
\begin{align} \label{eq:HM}
	H = &-t\sum_{\langle ij\rangle,\sigma}
	(c^\dagger_{i\sigma}c^{\phantom\dagger}_{j\sigma}+h.c.) + U \sum_i
	n_{i\uparrow}n_{i\downarrow}\nonumber\\ &+ V \sum_{\langle
	ij\rangle,\sigma\sigma'} n_{i\sigma}n_{j\sigma'}
\end{align}
has the usual near neighbor hopping $t$, on-site Coulomb repulsion $U$ and an
additional near neighbor Coulomb repulsion $V$. Weak coupling studies for $U
\ll W$ of this model, where $W=8$ is the bandwidth, have found that $d$-wave
pairing and $T_c$ are generally suppressed by $V$, but superconductivity
survives provided that $V$ is not larger than $\sim U^2/W$
\cite{Kivelson2012,Onari}. Variational Monte Carlo calculations of the model
in Eq.~\eqref{eq:HM} with an additional near neighbor exchange interaction $J$
have found that the on-site $U$ effectively enhances the $d$-wave pairing
interaction $J$, while suppressing the opposing effects of $V$, so that for
$U=10$, $d$-wave pairing is preserved up to $V=4J$ \cite{Plekhanov03}. Density
matrix renormalization group studies of a striped $t-J-V$ model, the strong
coupling $U\gg W$ limit of Eq.~\eqref{eq:HM}, have demonstrated that a
non-local $V$ can even lead to an enhancement of superconducting pair-field
correlations by inducing transverse stripe fluctuations \cite{Arrigoni02}. In
recent work using cellular dynamical mean field theory (CDMFT), S\'en\'echal
{\it et al.} found that $d$-wave pairing at zero temperature is preserved at
strong coupling even for $V\gg J$ as long as $V \lesssim U/2$
\cite{Tremblay2013}. An extension of this work to finite temperatures found
that at weak doping a finite $V$ can even lead to an increase in $T_c$, while
at large doping $V$ reduces $T_c$ \cite{Tremblay2016}. Based on a detailed
analysis of the frequency dependence of the gap function, the authors argued
that $V$ gives rise to a low frequency pairing contribution through an
increase in the effective exchange interaction $J = 4t^2/(U-V)$, while at high
frequencies, $V$ suppresses pairing. These studies thus concluded that
retardation plays an important role.

Here we use a similar cluster dynamical mean field treatment to examine the
$V$-dependence of $T_c$ and the dynamics of the pairing interaction in this
model. While the previous CDMFT calculations were carried out inside the
$d$-wave superconducting phase of model~\eqref{eq:HM}, our work directly
examines the dynamics of the pairing interaction in the normal state, and thus
provides new and complementary insight. In particular, we use the dynamical
cluster approximation (DCA) \cite{Hettler98,MaierRMP} with a continuous time
auxilary field (CT-AUX) quantum Monte Carlo (QMC) cluster solver \cite{GullCTAUX} to
perform numerical calculations of the model in Eq.~\eqref{eq:HM}.

The DCA maps the bulk lattice problem onto a finite size cluster of size $N_c$
and uses coarse-graining to retain the remaining degrees of freedom as a
mean-field that is coupled to the cluster degrees of freedom
\cite{Hettler98,MaierRMP}. The intra-cluster contribution of the interaction $V$
is treated exactly with QMC, while the inter-cluster terms may be treated
with an additional bosonic dynamic mean-field \cite{Maier05,Haule07} similar to
the extended dynamical mean-field theory \cite{Smith00}. Here, instead, we use
a Hartree approximation \cite{Tremblay2013}, which reduces to a shift in the
chemical potential in the absence of charge order \cite{Tremblay2016}. Due to
the neglect of dynamic inter-cluster effects of the interaction $V$, we do not
coarse-grain $V$ despite its non-locality.

For the small 2$\times$2 cluster we use, the sign problem of the underlying
CT-AUX QMC solver \cite{GullCTAUX,submatrix} is managable up to $V \sim U/2$
down to temperatures $T\sim T_c$. While the results obtained from this
2$\times$2 cluster should be regarded as mean-field results for the $d$-wave
$T_c$, the pairing dynamics is expected to be well described since temporal
fluctuations are fully retained through the inclusion of the dynamic mean
field. Larger clusters were recently considered in a DCA study of the
half-filled model, which does not have a sign problem \cite{Terletska17}. We
use $t=1$ as the unit of energy and set $U=7$. All results are for a filling
$\langle n\rangle = 0.9$.

In order to calculate $T_c$, we solve the Bethe-Salpeter equation (BSE) in the
normal state \cite{Maier06}
\begin{align} \label{eq:BSE}
	-\frac{T}{N_c}\sum_{\bK,\omega_n}
	&\Gamma^{pp}(\bK,\omega_n,\bK',\omega_{n'})
	\bar{\chi}_0^{pp}(\bK,\omega_{n'})\phi_\alpha(\bK',\omega_{n'})
	=\nonumber\\ &=\lambda_\alpha(T) \phi_\alpha(\bK,\omega_n)\,.
\end{align}  
Here $\Gamma^{pp}(\bK,\omega_n,\bK',\omega_{n'})$ is the irreducible
particle-particle vertex of the effective cluster problem with the cluster
momenta $\bK$ and Matsubara frequencies $\omega_n=(2n+1)\pi T$. The
coarse-grained bare particle-particle susceptibility
\begin{align}\label{eq:chipp}
	\bar{\chi}^{pp}(\bK,\omega_n) = \frac{N_c}{N}\sum_{{\bf k'}}G({\bf
	K+k'},\omega_n)G(-{\bf K-k'},-\omega_n)
\end{align}
is calculated from the single-particle Green's function $G({\bf k},\omega_n) =
[i\omega_n+\mu-\varepsilon_{\bf k}-\Sigma({\bf K},\omega_n)]^{-1}$ with $\mu$
the chemical potential, $\varepsilon_{\bf k}=-2t(\cos k_x+\cos k_y)$ the
dispersion and $\Sigma({\bf K},\omega_n)$ the cluster self-energy. Information
about the bulk lattice is retained through the ${\bf k'}$ sum
\cite{Jarrell01}, which runs over the $N/N_c$ momenta within a square patch
with $k_{x/y}\in [-\pi/2,\pi/2[$. At $T=T_c$ the leading eigenvalue of
Eq.~\eqref{eq:BSE} becomes 1 and the symmetry of the superconducting state is
given by the momentum and frequency dependence of $\phi({\bf K},\omega_n)$.
For all values of $V$ we consider, we find that the eigenvector corresponding
to the leading eigenvalue $\lambda_d$ has $d_{x^2-y^2}$-wave $\cos K_x - \cos
K_y$ structure.

\begin{figure}
\psfig{figure=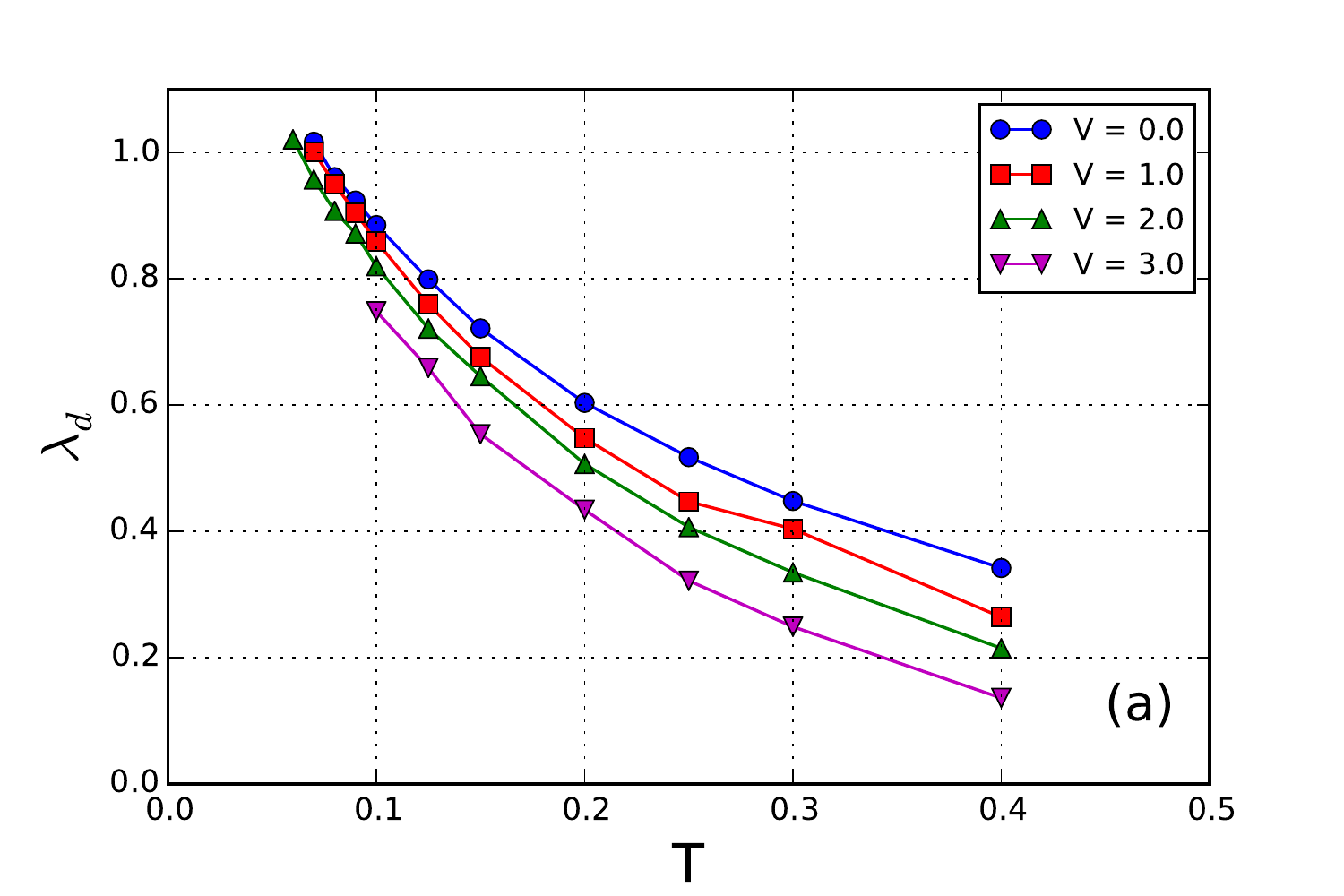,height=5.0cm,width=8.0cm,angle=0,clip}
\\
\psfig{figure=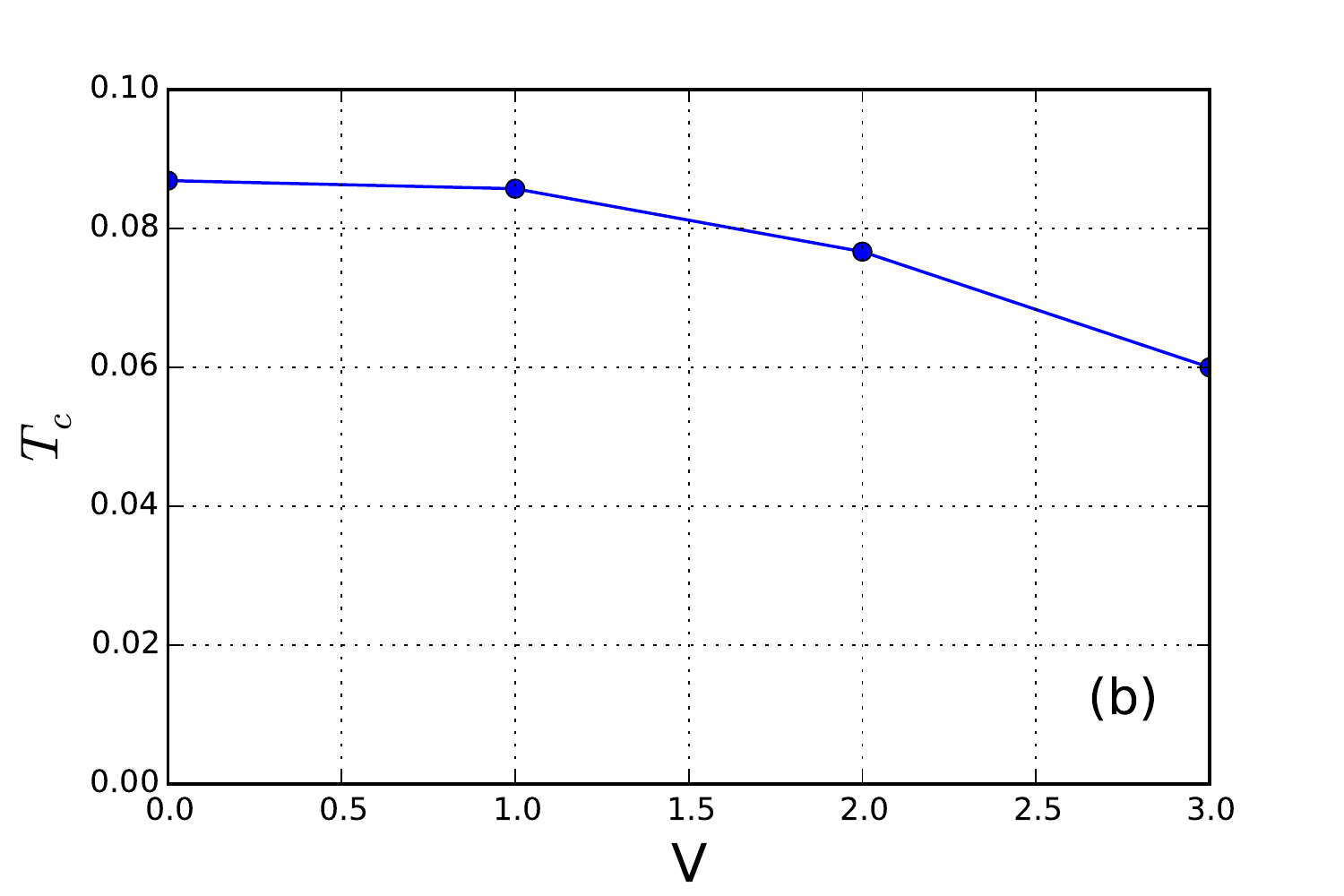,height=5.0cm,width=8.0cm,angle=0,clip} \\
\caption{(Color online) (a) Temperature dependence of the leading ($d_
{x^2-y^2}$-wave) eigenvalue $\lambda_ {d}(T)$ of the Bethe-Salpeter equation
in the particle-particle channel, Eq.~\eqref{eq:BSE} for the extended Hubbard
model in Eq.~\eqref{eq:HM} with $U=7$ and $\langle n\rangle=0.9$ for different
magnitudes of the near neighbor Coulomb repulsion $V$. (b) $d$-wave
superconducting transition temperature $T_{c}$ extracted from
$\lambda_d(T_c)=1$ as a function of $V$. $d$-wave pairing is only weakly
suppressed by the interaction $V$ as long as $V \lesssim U/2$.}
\label{lambda}
\end{figure}

Fig.~\ref{lambda}(a) shows the temperature dependence of the leading $d$-wave
eigenvalue $\lambda_d(T)$ of the BSE~\eqref{eq:BSE} for different magnitudes
of the nearest-neighbor repulsion $V$. As expected, finite $V$ leads to a
reduction of $\lambda_{d}(T)$ showing that $d$-wave pairing is weakened in the
presence of a nearest-neighbor repulsion.

From $\lambda_d(T_c)=1$ we can extract the $V$-dependence of $T_c$. For $V=3$,
where the QMC sign problem inhibits calculations down to $T_c$, we use a
polynomial fit of $\lambda_d(T)$ to extract $T_c$ from extrapolating to
$\lambda_d(T_c) =1$. As one sees from Fig.~\ref{lambda}(b), the $d$-wave $T_c$
is almost unchanged for $V=1$ and only slightly reduced by about 15\% for
$V=2$. The reduction becomes stronger for $V=3$ when $V$ approaches $U/2$.
This robustness of the $d$-wave pairing against a finite nearest-neighbor
repulsion is consistent with previous
studies~\cite{Kivelson2012,Tremblay2013,Tremblay2016}.

\begin{figure}
\psfig{figure=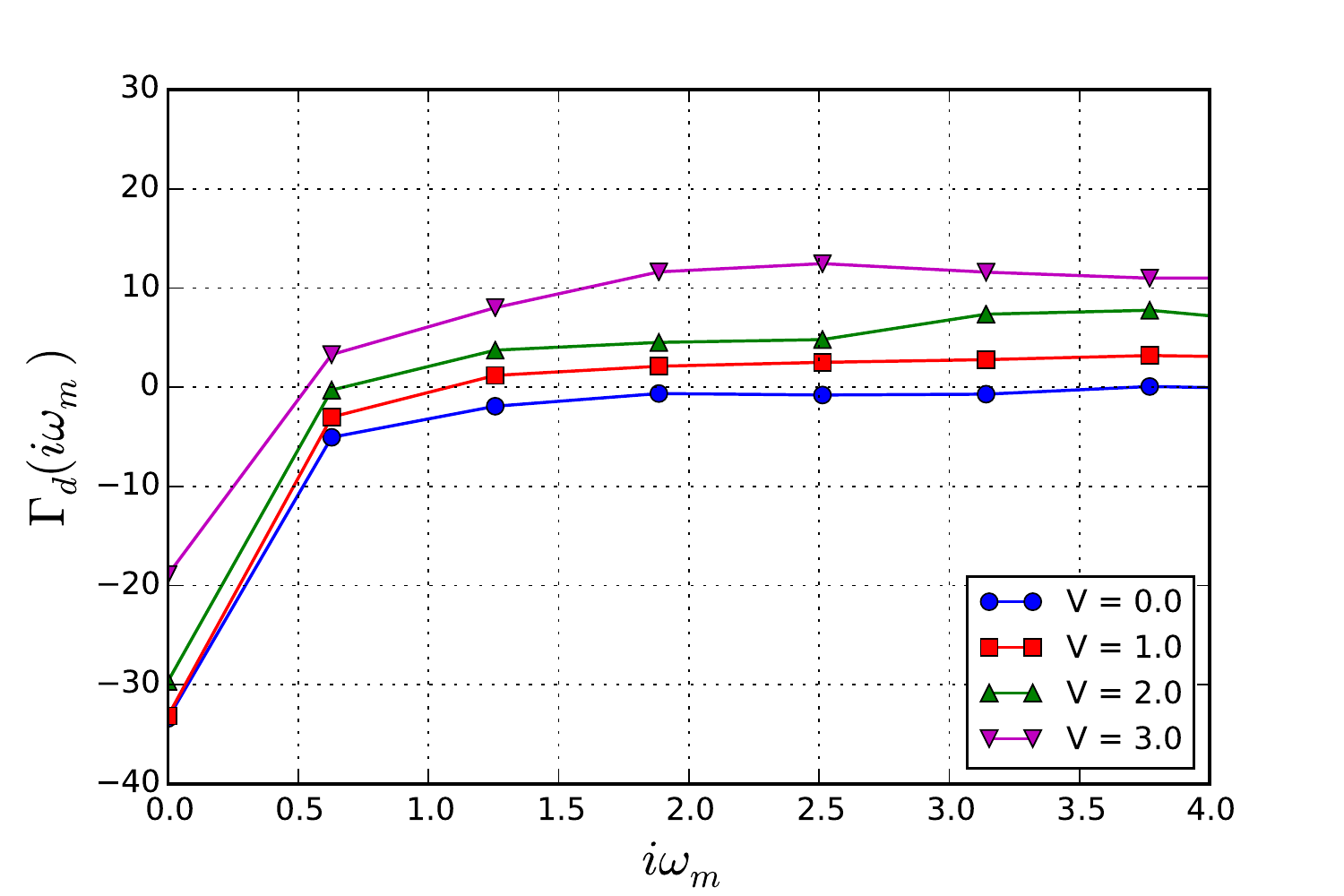,height=5.0cm,width=8.0cm,angle=0,clip} \\
\caption{(Color online) The $d$-wave projected irreducible particle-particle
vertex $\Gamma_{d}(\omega_{m})$ for different values of $V$. For
finite $V$, $\Gamma_d$ is attractive at low frequency but then turns repulsive
at higher frequency where it approaches $4V$.}
\label{Vd}
\end{figure}

In order to understand this resilience of $d$-wave pairing with respect to the
nearest neighbor Coulomb repulsion, we examine the dynamics of the pairing
interaction $\Gamma^{pp}({\bf K},\omega_n,{\bf K'},\omega_{n'})$ and the
leading $d$-wave eigenvector $\phi_d({\bf K},\omega_n)$. Fig.~\ref{Vd} shows a
plot of the frequency dependence of the $d$-wave projected pairing interaction
\begin{equation}\label{}
  \Gamma_{d}(\omega_m = \omega_n-\omega_{n'}) = \frac{\sum\limits_{
  \mathbf{K,K'}} g_d(
  \mathbf{K})
  \Gamma^{pp}(\mathbf{K},i\omega_{n} , \mathbf{K'},\omega_{n'})
  g_d(\mathbf{K'})}{\sum\limits_{\mathbf{K}} g_d^{2}(\mathbf{K})}\,.
\end{equation}
Here $g_d(
\mathbf{K})=\cos K_{x} - \cos K_{y}$ and we have set $\omega_{n'}=\pi T$ and
$T=0.1$. For $V=0$, $\Gamma_d(\omega_m)$ is negative (attractive)
for all frequencies and falls to zero at large $\omega_n$. For finite $V$,
$\Gamma_d(\omega_m)$ remains attractive at low frequencies, but then turns
positive (repulsive) at higher frequencies. This reflects the fact that at
high frequencies $\Gamma^{pp}(\bK,\omega_n,\bK',\omega_{n'}) \sim V(\bK -
\bK')$, where $V({\bf Q})$ is the Fourier-transform of the nearest neighbor
interaction $V$. For the 2$\times$2 cluster we have used here, one obtains
$\sum_{\bK,\bK'}g_d(\bK)V(\bK-\bK')g_d(\bK') / \sum_{\bK}g^2_d(\bK) = 4V$
consistent with the results in Fig.~\ref{Vd}. The dynamics of $\Gamma_d
(\omega_m)$ is reminiscent of the dynamics of the conventional
electron-phonon pairing interaction \cite{Schrieffer}, which is attractive at
low frequencies due to the effective electron-phonon attraction, and repulsive
at high frequencies due to the Coulomb repulsion. One also sees that $\Gamma_d
(\omega_m)$ becomes less attractive at low frequencies with increasing $V$.
This reduction even exceeds the frequency independent $4V$ repulsive
contribution, indicating that there is another repulsive and dynamic
contribution that further weakens the $d$-wave pairing interaction. We come
back to this point later when we examine the spin and charge susceptibilities.

\begin{figure}
\psfig{figure=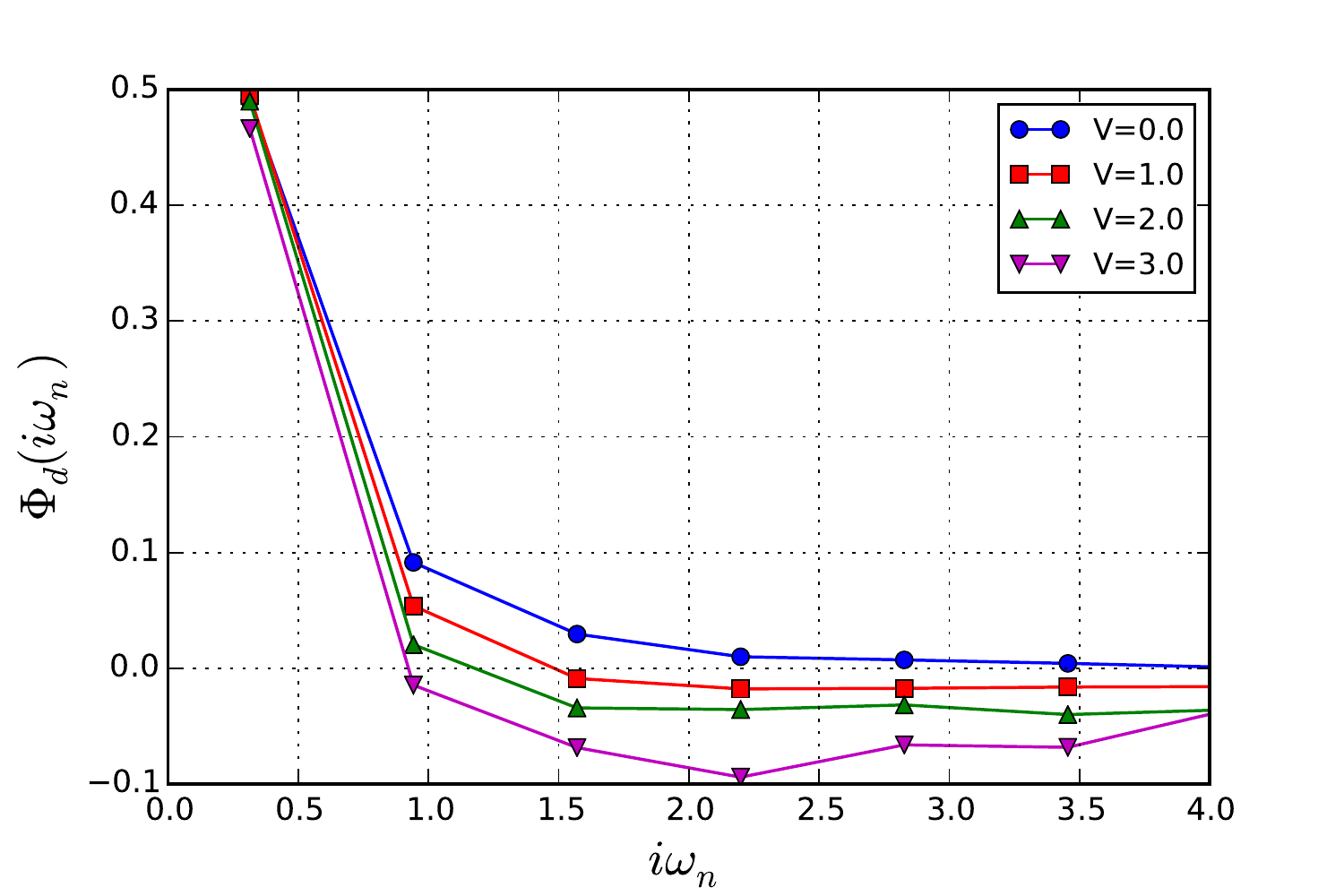,height=5.0cm,width=8.0cm,angle=0,clip} \\
\caption{(Color online) The frequency dependence of the leading $d$-wave
eigenvector $\phi_d({\bf K},\omega_n)$ of the Bethe-Salpeter equation
\eqref{eq:BSE} for ${\bf K}=(\pi,0)$ and $T/t=0.1$ for different values of
$V$. The sign change in $\phi_d ({\bf K},\omega_n)$ as a function of frequency
for finite $V$ minimizes the repulsive effect of $V$.}
\label{eigenvector}
\end{figure}

The dynamics of the pairing interaction is reflected in the frequency
dependence of the $d$-wave eigenvector $\phi_d({\bf K},\omega_n)$. This
quantity is plotted in Fig.~\ref{eigenvector} for ${\bf K}=(\pi,0)$ and
$T=0.1$ for different values of $V$. For $V=0$, $\phi_d((\pi,0),\omega_n)$
falls to zero on a characteristic frequency scale. As previously found in
Refs.~\cite{Maier06,Maier06PRB}, this scale is determined by the spin $S=1$
particle-hole continuum which for large $U$ is several times $J=4t^2/U$. For
finite $V$, the eigenvector changes sign and becomes negative at higher
frequencies. This sign change mirrors the sign change in $\Gamma_d(\omega_n)$.
Just as $\phi_d({\bf K},\omega_n)$ changes sign in ${\bf K}$-space reflecting
the repulsive nature of the pairing interaction at large momentum transfer
\cite{Maier06,ScalapinoRMP}, $\phi_d({\bf K},\omega_n)$ also changes sign in
frequency to adapt to the repulsive tail of the pairing interaction due to the
Coulomb $V$ at high frequencies. Thus, just as in the electron-phonon case,
retardation is important in preserving the attractive nature of the pairing
interaction in the presence of $V$.

\begin{figure}
\psfig{figure=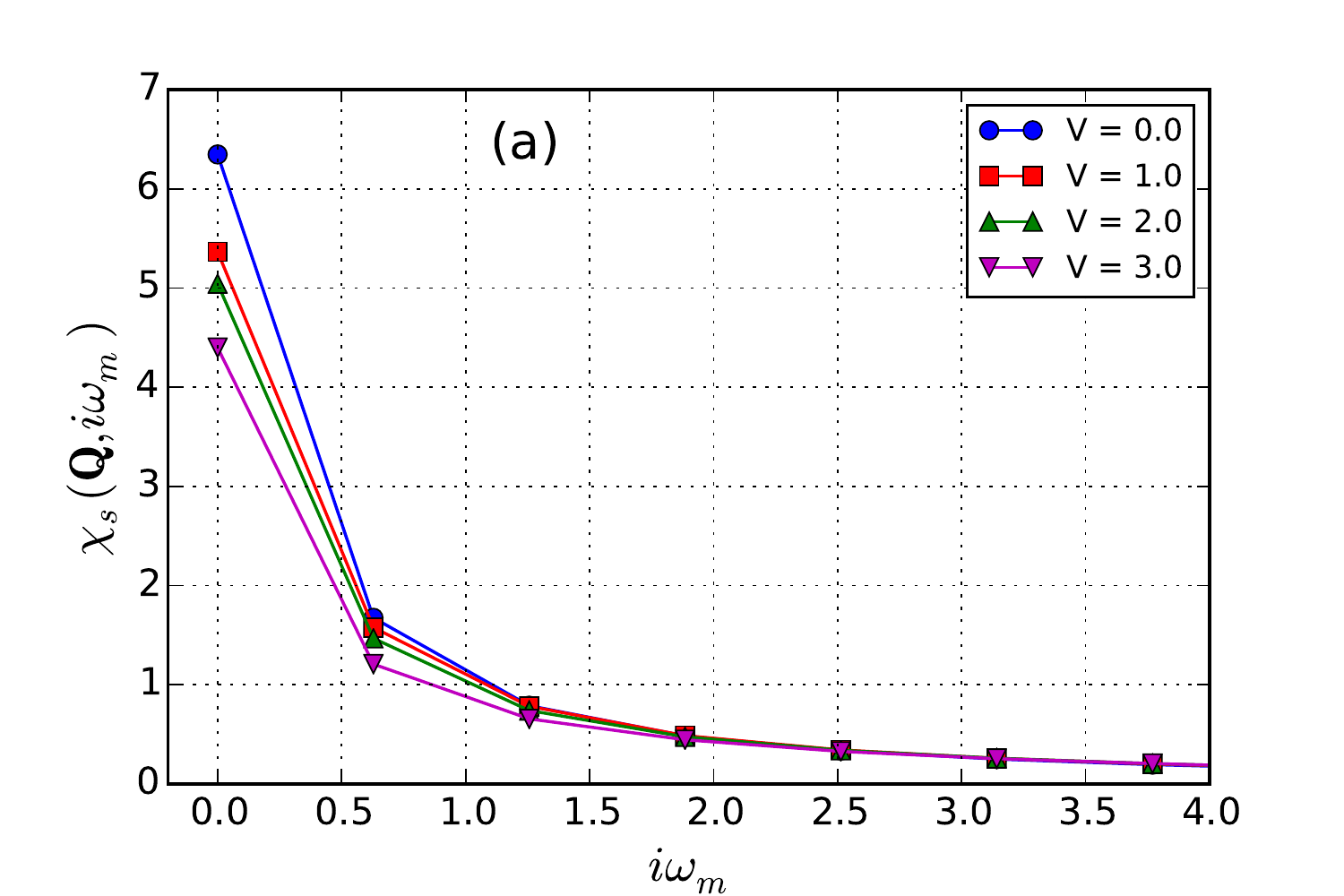,height=5.0cm,width=8.0cm,angle=0,clip} \\
\psfig{figure=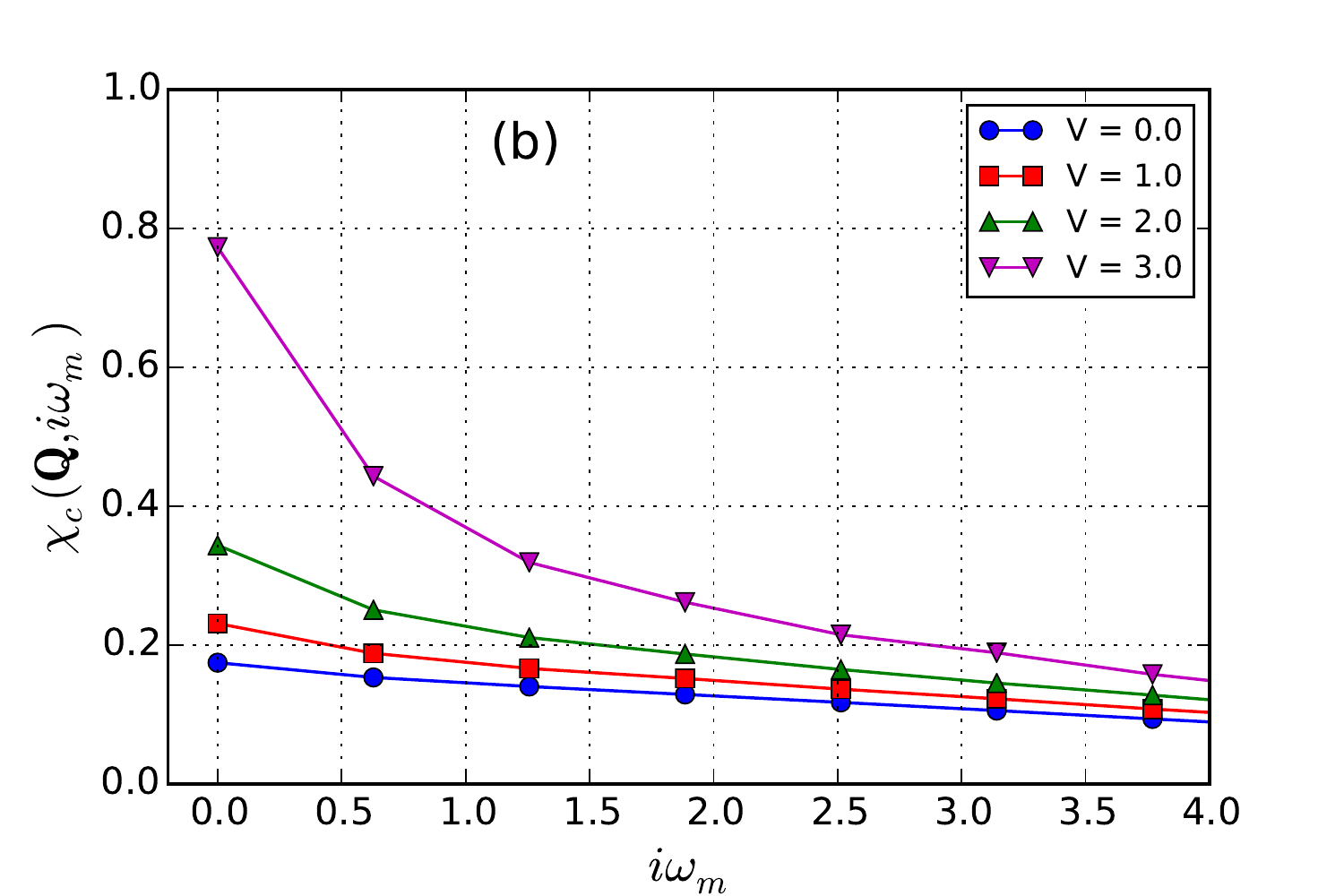,height=5.0cm,width=8.0cm,angle
=0,clip} \\
\psfig{figure=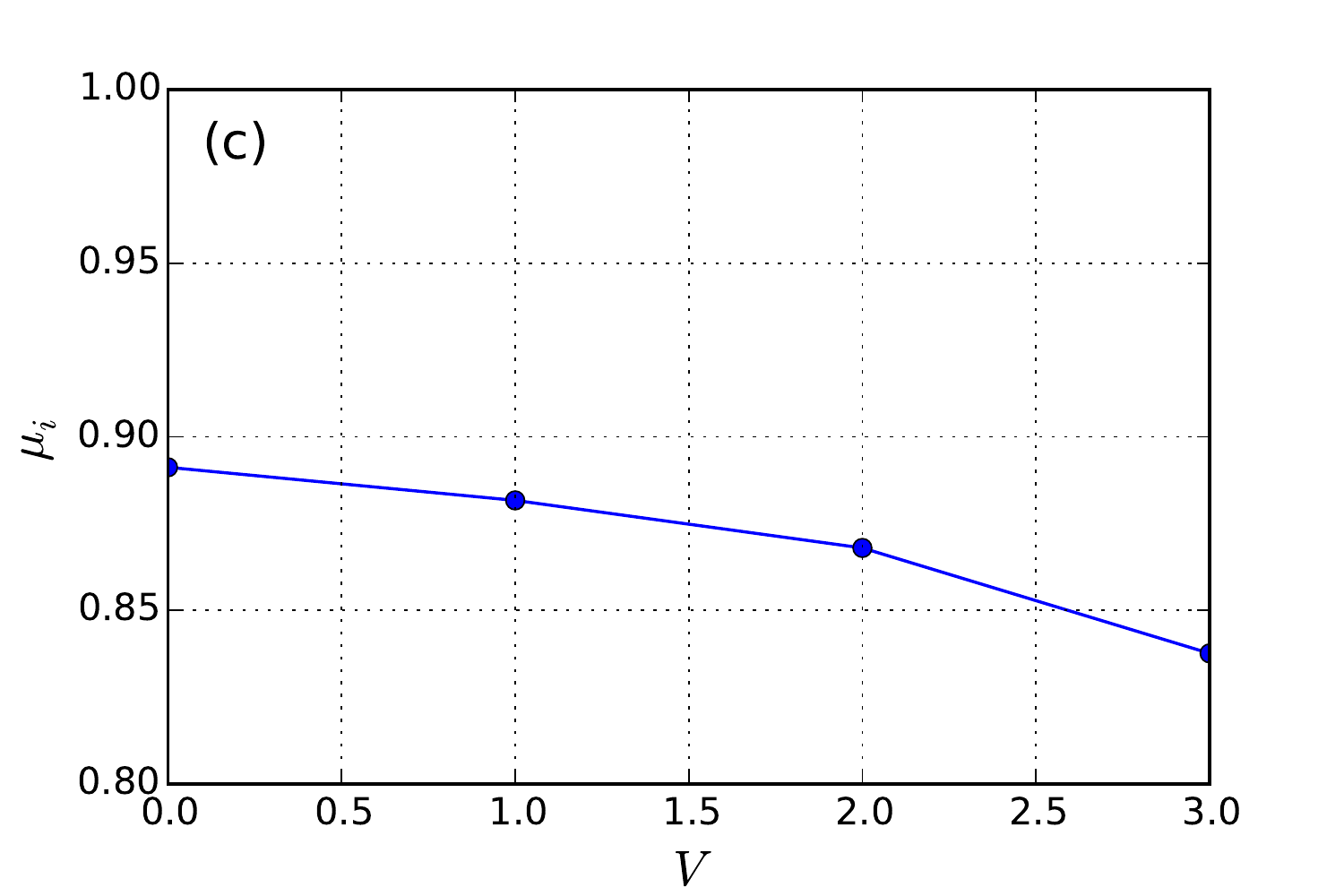,height=5.0cm,width=8.0cm,angle=0,clip} \\
\caption{(Color online) Frequency dependence of the spin and charge
susceptibilities, $\chi_s({\bf Q},\omega_m)$ in (a) and $\chi_c({\bf
Q},\omega_m)$ in (b), respectively, for $\mathbf{Q}=(\pi,\pi)$ and $T=0.1$ for
different values of $V$. (c) The local magnetic moment $\mu_i =
\sqrt{\langle (n_{i\uparrow} - n_{i\downarrow})^2\rangle}$ for $T=0.1$
as a function of $V$. With increasing $V$, charge fluctuations become
stronger while spin fluctuations are weakened through a reduction of the
local magnetic moment.}
\label{chi}
\end{figure}

We have also calculated the cluster dynamic spin (s) and charge (c)
susceptibilities
\begin{equation}\label{}
\begin{split}
  \chi_{s/c}(\mathbf{Q},\omega_{m}) = & \frac{1}{N_{c}} \sum\limits_{ij}
  e^{i\mathbf{Q} \cdot ({\bf R}_{i}-{\bf R}_{j})} e^{-i\omega_{m} \tau} \\ &
  \int_{0}^{\beta} d\tau \langle  [n_{i\uparrow}(\tau) \mp n_{i\downarrow}
  (\tau)] [n_{j\uparrow}(0) \mp n_{j\downarrow}(0)] \rangle\,.
\end{split}
\end{equation}
Here $\omega_m=2m\pi T$ are the bosonic Matsubara frequencies. The frequency
dependence of the spin- and charge-susceptibilities is shown in Figs.~\ref{chi}(a)
and (b) for ${\bf Q}=(\pi,\pi)$. As $V$ increases, $\chi_s({\bf Q},\omega_m)$
decreases at low frequencies, while $\chi_c({\bf Q},\omega_m)$ increases. The
rise in the charge susceptibility reflects the increasing tendency of the system
to form a $(\pi,\pi)$ charge density wave ordered state \cite{Kivelson2012}.

The $V$-dependence of the spin susceptibility is more difficult to understand.
Based on a strong-coupling picture, Reymbout {\it et al.} \cite{Tremblay2016}
have argued that a finite $V$ increases the exchange coupling $J=4t^2/(U-V)$
and thus the magnetic pairing mechanism. Our results for
$\chi_s((\pi,\pi),\omega_m)$, however, are not in line with this picture,
since one would would expect $\chi_s( (\pi,\pi),\omega_m=0)$ to increase with
$J$ and thus $V$. Rather, the decrease we observe can be understood from the
increase in the charge fluctuations. As shown in Fig.~4(c), these give rise to a
decrease in the local magnetic moment $\mu_i =
\sqrt{\langle (n_{i\uparrow} - n_{i\downarrow})^2\rangle}$, which leads to a
suppression of the spin fluctuations. The destructive effects of the
increasing charge fluctuations are thus two-fold: As shown previously,
$d$-wave pairing in the Hubbard model is mediated by a repulsive (positive)
pairing interaction in momentum space that increases with momentum transfer
and which reflects the momentum structure and dynamics of the spin
susceptibility \cite{Maier06,Maier06PRB}. Since charge fluctuations contribute
negatively to $\Gamma^{pp}(\bK,\omega_n,\bK',\omega_ {n'})$ \cite{Maier06},
large momentum charge fluctuations decrease the $d$-wave pairing interaction.
In addition, through their suppression of the local magnetic moment, they
further weaken the large momentum spin-fluctuations as seen in
Fig.~\ref{chi}(a).

To summarize, we have used dynamic cluster quantum Monte Carlo calculations of
an extended Hubbard model to study $d$-wave superconductivity when a near
neighbor Coulomb repulsion $V$ is present in addition to the on-site Coulomb
repulsion $U$. Consistent with previous studies, we find that $d$-wave pairing
and $T_c$ are only weakly suppressed by $V$ and remain stable as long as $V$
does not exceed $U/2$. The $d$-wave pairing interaction is attractive at low
frequencies and repulsive at high frequencies due to the repulsive effect
of $V$ on $d$-wave pairing. Reflecting this sign change, the $d$-wave
eigenfunction of the particle-particle Bethe-Salpeter equation, $\phi_d
(\bK,\omega_n)$, also changes sign as a function of frequency, similar to the
case of electron-phonon mediated pairing, thus reducing the repulsive effect of
the Coulomb interaction $V$. This demonstrates that retardation
plays an important role in stabilizing $d$-wave pairing in the presence of $V$.
A further analysis of the spin and charge susceptibilities shows that
$(\pi,\pi)$ charge-fluctuations become strong when $V$ approaches $U/2$. This
leads to a more rapid suppression of $d$-wave pairing and $T_c$ through both a
reduction of the $(\pi,\pi)$ spin fluctuations and a direct negative
contribution to the $d$-wave pairing interaction. 

\section*{Acknowledgments}

We acknowledge useful discussions with D.J. Scalapino. TAM was supported by the
U.S. Department of Energy, Office of Basic Energy Sciences, Materials Sciences and Engineering Division.


\bibliography{CurroBibliography}

\end{document}